\documentclass[onecolumn]{aa} 
\usepackage{graphicx}
\usepackage{txfonts}
\newcommand{\vphi}{\varphi}
\newcommand{\vpsi}{\varpi}
\def\beq{\begin{equation}}
\def\eeq{\end{equation}}
\def\beqar{\begin{eqnarray}}
\def\eeqar{\end{eqnarray}}

\newcommand{\OOmega}{{\cal A}}

\newcommand{\R}{{R_{\sun}}}

\newcommand{\mA}{{{S}}}
\newcommand{\mB}{{{M}}}

\newcommand{\meta}{{\overline{\eta}}}
\newcommand{\mnu}{{\overline{\nu}}}

\newcommand{\mN}{{\overline{N}}}

\newcommand{\mrho}{\overline{\rho}}

\newcommand{\ha}{\hat{a}}

\newcommand{\bk}{{\bf k}}
\newcommand{\p}{{\partial}}

\newcommand{\hf}{{\hat{ f}}}

\newcommand{\gapprox}{\lower.4ex\hbox{$\;\buildrel
>\over{\scriptstyle\sim}\;$}}
\newcommand{\lapprox}{\lower.4ex\hbox{$\;\buildrel
<\over{\scriptstyle\sim}\;$}}

\begin{document}
   \title{On a long-term dynamics of the magnetised solar tachocline}


   \author{Eun-jin Kim and Nicolas Leprovost
          }

   \offprints{E. Kim}

   \institute{Department of Applied Mathematics, University of Sheffield,
              Sheffield, S3 7RH, UK\\
              \email{e.kim@sheffield.ac.uk}
             }

   \date{Received 7 September, accepted, 2006}

  \abstract
   {}
{We investigate the confinement and long-term dynamics of the
magnetised solar tachocline.}
   {Starting from first principles, we derive
the values of turbulent transport coefficients
in the magnetised solar tachocline
and then explore the implications
for the confinement and long-term dynamics of the tachocline.} 
{For reasonable parameter values, the turbulent eddy viscosity is 
found to be negative, with turbulence enhancing the radial shear 
in the tachocline. Both magnetic diffusivity and thermal diffusivity are 
severely quenched, with the values much smaller than the magnitude of 
the eddy viscosity.  The effect of the meridional circulation on 
momentum transport via the hyperviscosity becomes important when
the radial shear becomes large (larger than the presently inferred 
value) due to the negative viscosity. The results imply that the 
tachocline develops too strong radial shear to be a stationary 
Hartmann layer. In the limit of a strong radiative damping
where the turbulence is active on very small scales ($< 10^{-4} \R$),
the eddy viscosity can become positive although
its effect is likely to be dominated by the hyperviscosity.
In comparison with the momentum transport, the transport of
magnetic field, heat, and passive particles is more
severely quenched. The results imply that the tachocline of thickness 
is larger than $O(10^{-2} \R)$, independent of the strength 
of magnetic fields. In addition, the momentum transport is much more 
efficient than the particle mixing in the tachocline, consistent 
with the observations.}
   {}
                                                                                
   
\titlerunning{Magnetised solar tachocline}                                                                             
     \maketitle


\section{Introduction}

A consistent theory of transport in the solar interior (in particular,
the tachocline) is
essential to the understanding of the evolution of solar rotation
and magnetic fields and the distribution
of chemical species. While turbulence is assumed to be absent in the
interior in the standard solar model (Stix 1989), 
observations (e.g., Pinsonneault
et al.  1989) and numerical
simulations (e.g., R\"udiger \& Kitchatinov 1996)
suggest that transport in this region --- although not so 
fast as turbulent transport
(e.g., such as in the convection zone) ---
should be faster than molecular processes to be consistent
with the current
rotational profile and surface depletion of light elements.
Such a modest transport could be due to waves
via dissipative processes (e.g., radiative damping of gravity waves).
Another interesting possibility, which has not received much attention,
is that turbulence is present in the interior due to a variety of
instabilities (e.g., see Spruit 1999 and references therein), 
but that the overall transport due to this
turbulence is considerably reduced as a result of turbulence
regulation. Our previous works (Kim 2005; Kim \& Leprovost 2006)
have shown that
stable stratification as well as shearing by the radial differential rotation
in the tachocline
can precisely do this as the excitation of gravity waves reduces the
stochasticity in turbulent flow while shearing enhances the overall
dissipation (see also Kim 2004).

The turbulent transport reduction can also be caused
by magnetic fields (e.g., Cattaneo \& Vainshtein 1991; 
Gruzinov \& Diamond 1994; Kim \& Dubrulle 2001; Kim 2006). 
In the tachocline, a strong toroidal
magnetic field of the strength $10^4 \sim 10^5$ G is
believed to be present, which
can easily be generated when a weak poloidal magnetic
field is sheared by  differential rotation in the tachocline.
A poloidal magnetic field here could be either of primordial origin
evolving on a long evolutionary time scale
(i.e., slow tachocline),
or generated by
dynamo process operating on fast time scale of the solar cycle
(i.e. fast tachocline) (see e.g., Gilman 2000; Petrovay 2003). Thus,
magnetic fields can potentially play a crucial role in the transport
of momentum, chemical species, and magnetic flux
on long and/or short time scales. In particular, on an
evolutionary time scale, the tachocline may be considered as
a boundary layer between the uniformly rotating radiative
interior and differentially rotating convection zone with
latitudinal variation 
(R\"udiger \& Kitchatinov 1996; Gough \& McIntyre 1998; MacGregor \& 
Charbonneau 1999),
and the dynamics of this boundary layer crucially depends
on the values of the effective magnetic diffusivity,
eddy viscosity, etc. 
The understanding of this boundary layer thus requires
the prediction of these turbulent transport coefficients,
derived from first principles. 
Furthermore, the momentum transport and chemical 
mixing across the tachocline play a crucial role
in the evolution of solar differential
rotation and the distribution of chemical species.
In particular, the present solar rotational profile and surface
depletion of light elements (lithium) (Schatzman 1993)
indicate that
the angular momentum transport must have been more efficient
than the particle mixing in the solar interior (e.g. Pinsonneault 
et al. 1989). 
This should be explained by a consistent theory of the momentum 
transport and chemical 
mixing in the tachocline, rather than invoking
a crude parameterization
as has often been done by previous authors.

The purpose of this paper is to provide a consistent theory
of turbulent transport in the magnetised tachocline and
then investigate its implications for a long-term dynamics
of the (slow) tachocline. Special attention is paid to
the elucidation of different effects on transport of shearing,
stable stratification, and magnetic fields,
identifying what is most likely to be the main
mechanism for turbulence regulation in the tachocline.
The remainder of the paper is organized as follows.
We elucidate the effects of gravity-Alfven waves on turbulent transport
in Sect. 2. In Sect. 3, we incorporate the effect of shear flow
given by the
radial differential rotation and provide the theoretical
predictions for turbulent coefficients in the stratified
magnetised tachocline. 	We elaborate on 
implications of the results for a long-term dynamics of the tachocline 
in Sect. 4. Section 5 is devoted to the discussion of the limit
of strong radiative damping.
Section 6 contains the conclusions and discussions.

\section{Turbulent transport in 2D MHD with density stratification}
We envision that turbulence in the tachocline is driven externally,
for instance, when plumes penetrating from the convection zone
randomly stir/perturb the region. In the
presence of density stratification and magnetic fields, this
random stirring will
excite both gravity and Alfven waves in the tachocline,
which will in turn increase the memory of fluid motion
which would otherwise be random and incoherent. As a result,
turbulent transport is expected to be reduced (e.g. see
Kim \& Leprovost 2006). 
%
In Kim \& Leprovost (2006), we have shown that the stable
stratification (gravity waves) leads to transport
property in the three dimensional hydrodynamic (3D HD)
turbulence very similar to that in the two dimensional
hydrodynamic (2D HD) turbulence (without the latitudinal
dependences) with a negative viscosity in both 2D and 3D. 
We recall that without stratification
the eddy viscosity is negative in 2D HD case while it is positive 
in 3D HD case. Magnetic fields also tend to make the property
of 2D turbulence more like that of 3D turbulence, leading
to a positive viscosity even in 2D (Kim \& Dubrulle 2001).
Therefore, we model our stratified magnetized tachocline by a 2D 
incompressible fluid in a local cartesian coordinates $x$ and $y$.
Here, $x$ and $y$
represent radial and azimuthal directions, respectively.
We consider a uniform toroidal magnetic field
${\bf B} = B_0 {\hat y} = (0, -\p_x a_0, 0)$
and represent the differential rotation by a large-scale
shear flow
${\bf U}_0 = U_0({\bf x}) {\hat y}$ in parallel to the toroidal
magnetic field.
Note that a similar cartesian 2D model for the tachocline
was adopted in Kim \& MacGregor (2001).
For simplicity, we adopt the Boussinesq approximation to capture the effect of
density stratification and the quasi-linear analysis by assuming
that total mass density $\rho = \rho_0 + \rho'$, particle
density for chemical species $n=n_0+n'$, vorticity $\omega = \omega' = \omega_z'
= \p_x v_y-\p_y v_x$, and
magnetic potential $a= a_0 + a'$ consist of mean and fluctuating
components, denoted by subscript `$0$' and prime, respectively.
The governing equations
for the fluctuations $\omega'$, $a'$, and $\rho_1 = \rho'/\mrho$
(${\mrho}$ is the mean constant mass density)
can be written as follows:
\begin{eqnarray}
(\p_t + U_0 \p_y )\omega'
&=& g \p_y \rho_1 - B_0 \p_y \nabla^2 a' + \nu \nabla^2 \omega' + {f}\,,
\label{eq1}\\
(\p_t + U_0 \p_y )a'
&=& -v_x  \p_x a_0  + \eta \nabla^2 a'\,,
\label{eq2}\\
(\p_t + U_0  \p_y) \rho_1
&=& {{{}} N^2 \over g} v_{x}+ \mu \nabla^2 \rho_1\,.
\label{eq3}
\end{eqnarray}
Here,
$\nu$, $\eta$ and $\mu$ are molecular viscosity, Ohmic
diffusivity and thermal diffusivity, respectively;
${f}$ in Eq. (\ref{eq1}) is the small-scale forcing
driving turbulence;
$N=\sqrt{-g(\p_x \rho_0 + {\mrho} g/c_s^2)/
{\overline {\rho}}}$ is the Brunt-V\"ais\"al\"a frequency;
$c_s$ is the sound speed; $\rho_0=\rho_0(x)$ and ${\mrho}$
are the mean
background and constant mass densities, respectively.
Note that the typical values of $\nu$, $\mu$, $\eta$, and $N$
in the tachocline are
$10^2$ cm$^2$s$^{{-1}}$,
$10^7$ cm$^2$s$^{{-1}}$, 
$10^4$ cm$^2$s$^{{-1}}$, and
$10^{-3}$ s$^{{-1}}$, 
respectively.

To elucidate the role of magnetic fields and stable stratification
in transport,
it is illuminating to examine their effect on the diffusion
of magnetic flux. To this end, we ignore the  large-scale
shear flow $U_0$ and forcing ${f}$ and
recast Eqs. (\ref{eq1})--(\ref{eq3}) as follows:
\begin{eqnarray}
\p_t {\tilde \omega}
&=& i g k_y {\tilde \rho_1}  +i k_y B_0  k^2
{\tilde a} -\nu k^2 {\tilde \omega}\,,
\label{eq01}\\
\p_t {\tilde a}
&=& i B_0 k_y {\tilde \omega}/k^2 - \eta k^2 {\tilde a}\,,
\label{eq02}\\
\p_t  {\tilde \rho_1}
&=& {{{}} N^2 \over g} {ik_y {\tilde \omega} \over k^2}
- \mu k^2 {\tilde \rho_1}\,.
\label{eq03}
\end{eqnarray}
Here, ${\tilde \omega}$, ${\tilde a}$ and ${\tilde \rho_1}$
are the Fourier transform of $\omega'$, $a'$ and $\rho_1$.
Equations (\ref{eq02}) and (\ref{eq03}) give us
$\p_t {\tilde \rho_1}/\p_t {\tilde a}
\sim {\tilde \rho_1}/{\tilde a}
\simeq N^2/g B_0$ for small dissipation, which can then
be used in Eq. (\ref{eq01}) to obtain
\begin{equation}
\p_t {\tilde \omega}
\simeq i k_y  B_0 k^2  \left[ {N^2 \over k^2 B_0^2}+1\right] {\tilde a} \,,
\label{eq4}
\end{equation}
By using Eq. (\ref{eq4}),
we can compute the magnetic flux as
\begin{eqnarray}
\Gamma_x &= &\langle v_x a' \rangle
\nonumber \\
&=& \int d^2 k d t \langle \p_t {\tilde v}_x ({\bf k}) {\tilde a}(-{\bf k})
 + {\tilde v}_x ({\bf k}) \p_t {\tilde a}(-{\bf k}) \rangle
\nonumber \\
&\simeq & {\tau B_0 \over 2} \left [ \langle v^2 \rangle - \alpha \langle b^2 \rangle
\right]\,.
\label{eq5}
\end{eqnarray}
Here, ${\bf b} = \nabla \times a' {\hat z}$,
$\alpha = 1 + N^2/k^2 B_0^2$, and
$\int dt \sim \tau$ is used where $\tau$ is
the characteristic time scale of turbulence
(see, e.g., Gruzinov \& Diamond 1994).
For stationary fluctuations, the flux $\Gamma_x$ is related to
magnetic energy $\langle b^2 \rangle$ as $B_0 \Gamma_x = \eta
\langle b^2 \rangle$,  thereby giving us $\eta_T = \Gamma_x/B_0$ as:
\begin{equation}
\eta_T \simeq {\eta_T^0 \over 1 + \alpha {\eta_T^0 B_0^2 \over \eta
\langle v^2 \rangle}}\,,
\label{eq6}
\end{equation}
where $\eta_T^0 = \tau \langle v^2 \rangle /2$ is the turbulent
diffusivity in the absence of magnetic field ($B_0 \to 0$) and
stratification ($N \to 0$). 
$\eta_T$ in Eq. (\ref{eq6}) immediately shows that magnetic
diffusivity is severely quenched for large $R_m = \eta_T^{0}/\eta$
as either the magnetic field or density stratification becomes strong.
This is because the excitation of waves
(Alfven or gravity waves) increases the memory of turbulent
eddies, reducing their stochasticity which is essential for
turbulent transport.
Note that Eq. (\ref{eq6}) recovers the 2D MHD result as $N \to 0$
(e.g. Gruzinov \& Diamond 1994).

To elucidate the effect of radiative damping, 
it is instructive to consider the limit of strong radiative damping.
In that case, Eq. (\ref{eq03}) can be approximated
as ${{{}} N^2} {ik_y
{\tilde \omega} /g k^2} \sim \mu k^2 {\tilde \rho}_1$.
A similar
analysis then gives us the magnetic diffusivity
\begin{equation}
\eta_T \simeq {\eta_T^0 \over  \lambda + {\eta_T^0 B_0^2 \over \eta
\langle v^2 \rangle}}\,,
\label{eq7}
\end{equation}
where $\lambda = 1+ \tau k_y^2 N^2/\mu k^4$.
The contribution from the gravity waves to the transport reduction
(the denominator of $\eta_T$) in Eq.\ (\ref{eq7}) is
$\tau k_y^2 N^2 / \mu k^4$, which should be compared with 
$\tau N^2 /\eta k^2$ in Eq.\ (\ref{eq6}). The ratio
of the two is thus $\eta/\mu$, which becomes very small
for strong radiative damping (i.e., for large $\mu$). That is,
magnetic diffusivity in Eq. (\ref{eq7}) is less reduced by
stratification in the case of a strong radiative damping,
as compared to that in the case of weak damping 
[Eq. (\ref{eq6})]. This is simply because
the radiative damping weakens buoyancy effect.
In the extreme limit of $\mu \to \infty$, the effect of
stratification disappears with $\lambda \to 1$ in Eq. (\ref{eq7}),
recovering 2D MHD result in the absence of stratification.

\section{Consistent theory with the radial differential rotation}
In Sect. 2, we have shown that both stable stratification and magnetic fields
can severely quench the transport of magnetic field. In
this section, we present
a consistent theory of turbulent transport in the magnetised tachocline,
by incorporating a background shear flow $U_0 {\hat y}$ provided by the radial
differential rotation. 
The latitudinal differential rotation is neglected compared
to radial differential rotation since it is weaker in the tachocline
due to thin tachocline thickness ($<0.03 \sim 0.05$ of the solar radius)
[see Leprovost \& Kim (2006) for the dynamics and the effect of latitudinal
differential rotation].
This shear flow plays a crucial role in
regulating turbulence and turbulent transport by shearing 
[e.g., see Kim (2005)]. For simplicity, we assume a linear shear flow 
$U_0 {\hat y}
= -x \OOmega {\hat y}$, where $\OOmega = |\p_x U_0|>0$ is the 
radial shear (or shearing
rate), which is assumed to be positive without loss of generality. 
In order to incorporate the shearing effect non-perturbatively, we employ
the special Fourier transform for the fluctuating quantities $\phi'$:
\begin{equation}
\phi'({\bf x},t) = {1\over (2 \pi)^3} \int d^3 k
\tilde{\phi}({\bf k},t) \exp{\{i(k_x(t) x + k_y y + k_z z)\}}\,,
\label{eq9}
\end{equation}
where $k_x(t)$ linearly increases in time due to shearing as
\begin{equation}
k_x(t) = k_x(0) + k_y \OOmega t\,.
\label{eq10}
\end{equation}
Here, $\OOmega = |\p_x U_0|$ is again the shearing rate of the
shear flow.
In this section, we consider the case where the radiative damping
rate is small compared to the shearing rate. That is, we consider a strong
shear limit where 
$\xi_\mu = \mu k_y^2/ \OOmega \ll 1$. Here, $l_f=1/k_y$ is the
characteristic length scale of the forcing. 
Thus, $\xi_\mu$ ($\ll 1$) is a small parameter, representing
the strong shear limit. For typical solar parameters
$\mu \sim 10^7$ cm$^2$ s$^{-1}$ and
$\OOmega \sim 3 \times 10^{-6}$ s$^{-1}$, 
this is a good assumption
valid for  a broad range
of characteristic length scales of the forcing 
$l_f = 1/k_y \gapprox 10^6\sim 10^7$ cm ($\sim 10^{-4} \R$ where
$R_{\sun} \sim 5 \times 10^{10}$ cm is the solar radius). 
Note that the opposite limit of the strong radiative
damping is considered in Sect. 5. 

For $\xi_\mu \ll 1$, the coupled equations (\ref{eq1})--(\ref{eq3}) can
easily be combined to form the following equation for
${\hat a} = {\tilde a} \exp{[\mu (k_x^3/3 k_y \OOmega + k_y^2 t)/2]}$
\begin{equation}
\p_\tau[(1 +\tau^2)\p_\tau \ha] + {\gamma^2}
[(1+\tau^2) + \beta] \ha
\simeq  {i B_0 \over \OOmega^2 k_y} {\hf}(\tau)\,,
\label{eq11}
\end{equation}
to leading order in $\xi_\mu$.
Here, $\tau = k_x/k_y = k_x(0)/k_y + \OOmega t$, $\gamma = |B_0 k_y/\OOmega|$,
$\beta = N^2/B_0^2 k_y^2 = N^2/\omega_A^2$, $\omega_A = |B_0 k_y|$
is the Alfven frequency of the mode $k_y$, and
${\hf} = {\tilde f} \exp{[\mu (k_x^3/3 k_y \OOmega + k_y^2 t)/2]}$.

The solution to Eq. (\ref{eq11}) in the limit of strong magnetic
field $\gamma \gg 1$ and stratification $N^2/\OOmega^2 \gg 1$ can
be found as
\begin{equation}
\ha (\tau) \simeq {i B_0 \over \OOmega^2 k_y \gamma}
\int d\tau_1 {\sin{\gamma[Q(\tau)-Q(\tau_1)]}\hf(\tau_1)
\over [(1+\tau^2) (1+\tau_1^2) (1+\tau^2 +\beta) (1+\tau_1^2+\beta)]^{1/4}}\,,
\label{sola}
\end{equation}
where $Q(\tau)= \int d\tau \sqrt{{1+\tau^2 + \beta \over 1+\tau^2}}$.
Note that the strong magnetic field limit $\gamma> 1$ holds
on a broad range of length scales of the forcing
$l_f \lapprox L_B = 10^{10}$ cm for $B_0 =10^4
\sim 10^5$ G in the tachocline.
To obtain the turbulent coefficients, we assume
that the forcing is homogeneous with a short correlation time $\tau_f$;
\begin{equation}
\langle {\tilde f} (\bk_1,t_1) {\tilde f} (\bk_2,t_2) \rangle
= \tau_f (2 \pi)^2 \delta (t_1-t_2) \delta(\bk_1+\bk_2)
{\phi} (\bk_2)\,.
\label{eq18}
\end{equation}
A long but straightforward algebra by using
Eqs. (\ref{eq1})-(\ref{eq4}), (\ref{sola}) and (\ref{eq18})
then gives us the eddy viscosity defined by
$\langle v_x v_y -b_x b_y \rangle = -\nu_T \p_x U_0 =\nu_T \OOmega$ (${\bf b} = \nabla
\times a' {\hat z}$):
\begin{eqnarray}
\nu_{T}
&=& {\tau_f\over  2 \OOmega^2} \int {d^2k \over (2 \pi)^2}
{\phi(\bk)\over k_y ^2} \left [-1+{1\over \sqrt{1+\beta}}\right]
\nonumber \\
&\sim&
{1\over \OOmega^2} 
\left [-1+{1\over \sqrt{1+\beta}}\right] {\cal F}
\,.
\label{eq29}
\end{eqnarray}
Here, $\phi({\bf k})$ is the power spectrum of the forcing
defined in Eq. (\ref{eq18}); 
${\cal F} \sim \int d^2 k \phi({\bf k})/k_y^2 \sim v_f^2/\tau_f$ 
is the strength of the forcing
with the characteristic velocity $v_f$ and correlation time  $\tau_f$;
$\beta = N^2/B_0^2 k_y^2 = N^2/\omega_A^2$, $\omega_A = |B_0 k_y|$;
$\OOmega = |\p_x U_0|$ is the radial shear.
What is the most remarkable about the eddy viscosity in Eq. (\ref{eq29}) is
that it is always negative regardless of the
relative strength of stratification to magnetic field (i.e., $\beta
= N^2/\omega_A^2$). In 2D MHD, the Maxwell stress
exactly cancels the Reynolds stress to leading order while
the incomplete cancellation between the two in the next
order gives
a small positive eddy viscosity ($\propto 1/B_0^2$)
(Kim \& Dubrulle 2001).
The cancellation of these leading order contributions in
2D unstratified MHD can easily be
checked in Eq. (\ref{eq29}) by putting $\beta = 0$, which
gives $\nu_T = 0$.
In contrast, in a stratified medium with $\beta \ne 0$,
the leading order cancellation in Eq. (\ref{eq29}) is not perfect,
giving a net negative eddy viscosity.
This can be shown to be due to the fact
that the exact equipartition
between the kinetic energy and magnetic energy (for pure
Alfven waves) is broken by buoyancy, with larger kinetic energy
than magnetic energy, driving a negative eddy viscosity.
A negative eddy viscosity, in a sharp contrast to a positive
eddy viscosity, signifies an anti-diffusive momentum transport
against the gradient of the shear flow. In other words, the
overall momentum transport due to turbulence accelerates the
mean flow, accentuating its gradient rather than eradicating it.
A similar tendency of the acceleration of the mean flow
in the magnetised tachocline can also be due to the direct momentum
deposition of gravity waves via radiative damping (Kim \& MacGregor 2001).

For clarity, we examine the
behavior of $\nu_T$ as a function of $\beta$.
For large $\beta$ (strong stratification/weak magnetic field),
$\nu_T \propto [-1 + |B_0 k_y/N|]/\OOmega^2$, showing
that the effect of magnetic fields tends to make
eddy viscosity positive. In the opposite limit of
small $\beta$ (weak stratification/strong magnetic field),
$\nu_T \propto -N^2/B_0^2 \OOmega^2$, whose
absolute magnitude $|\nu_T|$ is small compared
to 2D HD case. This again reflects
the tendency of magnetic fields making the eddy viscosity
less negative (i.e. more positive).
Note that for parameter values
typical of the tachocline, the cross-over scale $L_N$
from $\beta>1$ to $\beta<1$ is roughly
$L_N = 10^7 \sim 10^8$ cm.
To recapitulate, the result (\ref{eq29}) shows a
tendency of a negative eddy viscosity
in stratified medium despite the presence of strong
magnetic fields, suggesting that the turbulent
transport in the tachocline would amplify the shear
provided by the radial differential rotation
for reasonable values of parameters.
We emphasize that the negative eddy viscosity represents
the amplification of a large-scale shear flow
at the expense of small-scale turbulence.
The value of magnetic diffusivity, defined by
$\langle a' v_x \rangle = -\eta_T \p_x a_0$,
depends on whether
$\alpha = (\xi_\mu/3)(1+\beta)^{3/2}$ is larger or smaller
than unity. First, in the case $\alpha \ll 1$,
which is valid for either $\beta>1$ or $\beta<1$ on scales
$L<L_M = B_0^3 \OOmega/ \mu N^3\sim 10^8$ cm ($10^{11}$ cm)
for $B_0\sim 10^4$ G ($10^5 $ G), we can obtain
\begin{eqnarray}
\eta_{T}
&= &{\tau_f \over  2 B_0^2} \int {d^2k \over (2 \pi)^2}
{\phi(\bk)\over k_y ^4} \xi_\eta G_0 {1 \over \sqrt{1+\beta}}
\nonumber \\
&\sim&
 {1 \over   B_0^2k_y^2 } \xi_\eta^{2/3} \left({\eta\over \mu}\right)^{1/3}
 {1 \over \sqrt{1+\beta}} {\cal F} \,.
\label{eq30}
\end{eqnarray}
Here, $G_0 = (3/\xi_\mu)^{1/3}\Gamma(1/3)/3$;
$\xi_\mu = \mu k_y^2/\OOmega\ll 1$;
$\xi_\eta = \eta k_y^2/\OOmega\ll 1$;
$\Gamma(x)$ is the Gamma function;
${\cal F} \sim \int d^2 k \phi({\bf k})/k_y^2 \sim v_f^2/\tau_f$ 
is again the strength of the forcing
with the characteristic velocity $v_f$ and correlation time  $\tau_f$.
In comparison, for $(\beta \gg)$ $ \alpha \gg 1$, we obtain
\begin{eqnarray}
\eta_{T}
&= &{\tau_f \over  2 N^2} \int {d^2k \over (2 \pi)^2}
{\phi(\bk)\over k_y ^2} \xi_\eta G_1
\nonumber \\ 
&\sim&
 {1 \over  N^2} \xi_\eta^{1/3} \left({\eta\over \mu}\right)^{2/3}
{\cal F}
\,,
\label{eq31}
\end{eqnarray}
where $G_1 = (3/\xi_\mu)^{2/3}\Gamma(2/3)/3$.
In each case, the density (heat) diffusivity $\mu_T$
($\langle \rho' v_x \rangle = -\mu_T \p_x \rho_0$)
is given by
\begin{equation}
\mu_T \simeq {\mu\over \eta} \eta_T\,.
\label{eq32}
\end{equation}
(see also Kim 2006).

Equation (\ref{eq30}) shows that
the turbulent diffusion of magnetic field ($\eta_T$)
can severely be quenched by  a strong mean magnetic field and stratification,
proportional to $1/B_0^2$ for $\beta<1$ and to $1/B_0 N$ for $\beta>1$,
respectively.
It is worth  noting that $\eta_T \propto 1/B_0^2$ was 
observed in numerical simulation (e.g. Cattaneo and Vainshtein 1991)
of 2D unstratified MHD turbulence ($\beta=0$).
As the stratification becomes stronger for a fixed $B_0$ with the 
further increase in
$\beta$, the diffusion is now reduced as $1/N^2$ [see Eq.\ (\ref{eq31})].
It is important to note that in all cases,
magnetic diffusivity has a much smaller magnitude
than the eddy viscosity [in Eq. (\ref{eq29})],
with a small value of $\eta_T/|\nu_T|$.
Note that a similar tendency was also found in
the stably stratified shear turbulence
without magnetic fields (Kim \& Leprovost 2006).
Specifically, in the strong magnetic
field and weak stratification region with $\alpha<1$ and $\beta>1$,
$\eta_T/|\nu_T| \sim \xi_\eta^{2/3} (\eta/\mu)^{1/3} \OOmega^2 /N \omega_A
\ll 1$.
Furthermore, the heat diffusivity $\mu_T$ in Eq. (\ref{eq32}),
although larger than $\eta_T$ by a factor of $\mu/\eta = 10^3$,
is yet much smaller than the magnitude of $\nu_T$.
For instance, again when $\alpha<1$ and $\beta>1$,
this ratio becomes $\mu_T/|\nu_T| \sim
\xi_\mu^{2/3}  \OOmega^2 /N \omega_A \ll 1$ for typical
parameter values. These results have very interesting implications
for a long-term dynamics of the tachocline, as discussed in Sect. 4.

\section{Implications for a long-term dynamics of the tachocline}

The results of the present and previous paper (Kim \& Leprovost 2006;
Kim \& MacGregor 2001) suggest that the uniform rotation in the
radiative interior is very unlikely to be explained by hydrodynamical
means as the momentum transport in stratified medium accelerates
the mean flow, sharpening the gradient of radial differential
rotation that has been created during the solar spin-down
(see, however, Charbonnel \& Talon 2005).
Previous authors 
(R\"udiger \& Kitchatinov 1996; Gough \& McIntyre 1998; MacGregor \& 
Charbonneau 1999),
have however shown that a rather weak
poloidal magnetic field in the radiative interior
can eliminate the differential rotation, thereby
leading to a uniform rotation therein.
In this case, the tachocline can be envisioned
as a boundary layer where the generation of the toroidal magnetic
field by the shearing of the poloidal magnetic field 
(due to differential
rotation) is balanced by the diffusion of the toroidal magnetic
field while the dissipation of the radial differential
rotation is balanced by the azimuthal Lorentz force associated
with the large-scale toroidal and poloidal magnetic fields, i.e., Hartmann layer
(R\"udiger \& Kitchatinov 1996; Gough \& McIntyre 1998; MacGregor \& 
Charbonneau 1999).
By adopting the molecular values for viscosity,
magnetic diffusivity, and heat diffusivity
in the tachocline, 
these previous authors obtained estimates of the tachocline
thickness and the strength of the interior poloidal magnetic field.
In the case of the tachocline with residual turbulence,
for instance, driven externally (e.g. plume penetration) or internally
(e.g. via instability), the values of turbulent transport coefficients,
instead of molecular values, 
should be used in the analysis of the Hartmann layer.

For the clarity of the discussion, it is worth recalling that
the Hartmann layer is based on the configuration where
a poloidal magnetic field
is fully contained in the interior, without penetrating
into the convection zone above so that the latitudinal differential
rotation in the convection zone does not leave its footprint
into the radiative interior.
By representing the latitudinal coordinate by $z$, and
by denoting the poloidal and toroidal magnetic
fields by $B_z$ and $B_y=B_0$, respectively, the major
force balance for the toroidal magnetic field in the
tachocline $B_y$ and the mean shear flow $U_y$ due to
the differential rotation can roughly be expressed as
\begin{eqnarray}
\partial_t U_y & \sim& B_z \partial_z B_y + \nu_T \partial_{xx} U_y
- \lambda_T \partial_{xxxx} U_y
\,,
\label{eq38} \\
\partial_t B_y & \sim& B_z \partial_z U_y  + \eta_T \partial_{xx} B_y\,.
\label{eq39}
\end{eqnarray}
Here, $\lambda_T \sim \mu_T (\Omega/N)^2 (\R/\Lambda)^2$ is the hyperviscosity
due to the meridional circulation (see, e.g., Spiegel \& Zahn 1992); 
$\Omega \sim 3 \times 10^{-6}$ s$^{-1}$ and
$\R \sim 5 \times 10^{10}$ cm
are the average rotation rate and solar radius, respectively;
$\Lambda$ is a constant of order unity.
Similarly to a positive eddy viscosity,
a (positive) hyperviscosity acts to smooth out the gradient of (or damp)
a large-scale shear flow. 
%

%

While the values of turbulent transport ($\nu_T$, $\eta_T$, and $\lambda_T$) for the coupled system (\ref{eq38})-(\ref{eq39})
have conventionally been assumed to be positive,
the results in Sect. 3 show that the turbulent momentum
transport in the sheared stratified turbulent tachocline
is anti-diffusive with negative viscosity (i.e., $\nu_T<0$), 
accelerating the mean shear flow $U_y$ (i.e.,
radial differential rotation). 
The negative viscosity will amplifies the shear in the differential
rotation and consequently toroidal magnetic field since 
they are coupled via the Lorentz force (via $B_z$).
The crucial question is
then how the Hartmann layer is maintained with negative
viscosity which tends to become unstable due to $\nu_T<0$. 
Obviously, if the magnetic diffusivity $\eta_T$ is large
enough to overcome an unstable situation caused by the negative
viscosity, the coupled system can find a stable stationary
configuration. However, this is very unlikely since
the magnetic diffusion is pathetically small compared to
the magnitude of the eddy viscosity, as discussed in Sect. 3.
Could the hyperviscosity $\lambda_T \sim (\R/\Lambda)^2 (\Omega/N)^2 \mu_T$
due to the meridional circulation
then stabilize the system? 
To answer this question, we note that
the contribution from the hyperviscosity to
Eq. (\ref{eq38}) is of order $\lambda_T/h^4$
while the contribution
from the eddy viscosity is of order $\nu_T/h^2$,
where $h$ is of order of the tachocline thickness.
Thus, the ratio of the two is roughly $(\mu_T/\nu_T) (\R/h)^2 (\Omega/N)^2$
(e.g. $\sim (\mu k_y^2/\OOmega)^{2/3} (\OOmega^2/N \omega_A) (\Omega/N)^2
(\R/h)^2$ for $\alpha<1$ and $\beta>1$),
which can be shown to be small for the parameter values typical of the
present solar tachocline.
However, it is important to realize that this ratio ($\propto \OOmega^{4/3}$
for $\alpha<1$ and $\beta>1$, for instance)
becomes large as the shear $\OOmega$ increases. 
Therefore, it is plausible that
as the shear is amplified via the negative eddy viscosity,
the effect of the hyperviscosity becomes important and
eventually dominates the negative viscosity, possibly stabilizing the
system. In order for this to be the case,
the shear $\OOmega$ in the radial differential rotation
however has to be larger than what is observed today.
This can be checked by requiring $\lambda_T/h^2 \nu_T >1$.
For instance, in the case $\alpha<1$ and $\beta>1$, this
demands $\OOmega/N > (N/\mu k_y^2)^{1/2} (B_0 k_y /N)^{3/4}
(hN/\R\Omega)^{3/2}$,
where $l_f=1/k_y$ is the characteristic scale of the forcing.
Even if we take $B_0 \sim 10^4$ G and a thin tachocline 
$h \sim 10^{-3} \R$, $\OOmega > 0.1 N $
for $k_y \sim 10^{-7}$ cm$^{-1}$, and 
$\OOmega > 0.01 N $ for $k_y \sim 10^{-6}$ cm$^{-1}$,
which seem to be rather too large to be reasonable 
(recall $N \sim 10^{3}$ s$^{-1}$ and $\OOmega \sim 3 \times 10^{-6}$ s$^{-1}$
are the presently inferred values in the tachocline).

To make this argument more concrete, it is instructive to
examine the behavior of Eqs.\ (\ref{eq38})-(\ref{eq39})
in more detail.
To this end, we average
them over the space ($x$)
with the approximation $\p_z \sim i/\R$ and $\p_x \sim 1/h$
and use the values of $\nu_T$, $\eta_T$ and $\mu_T$ obtained
for $\alpha < 1$ 
in Sect. 3. 
Here, $h$ is again the tachocline thickness.
The resulting envelop equations for $\OOmega \sim U_y/\R$
and $B_y$, in a properly non-dimensonalized form,
are as follows:
\begin{eqnarray}
\p_t \mA  = i \mB  + [\mnu(\mA,\mB) - \beta \meta(\mA,\mB)] \mA \,,
\nonumber \\
\p_t \mB  = i \mA  - \meta (\mA,\mB) \mB \,.
\label{eq40}
\end{eqnarray}
Here, $\mA = \OOmega/\Omega$, $\mB = B_y/\Omega \R$,
$\mnu = \alpha[1-1/\sqrt{1+1/\mB^2}]/|\mA|^2$, and
$\meta = 1/\sqrt{\mB^4+\mB^2} |\mA|^{2/3}$. 
The parameter $\alpha$ appearing in $\mnu$ is
a rough measure of $|\nu_T/\eta_T|$, which was
shown to be very large in Sect. 3 because 
the magnetic diffusion is much smaller than
the magnitude of the eddy viscosity. 
Specifically, $\alpha \sim (\mu/\eta)^{1/3} (\OOmega/\eta k_y^2)^{2/3}
N^2/\OOmega^2 \gapprox O(10^8)$ obtained by using 
the reasonable parameter values in the tachocline 
$\OOmega \sim 3 \times 10^{-6}$ s$^{-1}$, 
$N \sim 3 \times 10^{-3}$ s$^{-1}$,
and $k_y \sim 10^{-8}$ cm$^{-1}$.
The parameter $\beta $ in Eq. (\ref{eq40}) 
represents a rough measure of the ratio of the effect of 
hyperviscosity to that of magnetic diffusivity, with
the value  
$\beta \sim (\R \Omega/h N)^2 (\mu/\eta) \sim
\lapprox O(10^3)$ for $h/\R > 10^{-3}$. 

We note that the values of
$\eta_T$, $\nu_T$ and $\mu_T$ are all proportional
to the intensity of the forcing ${\cal F}$, which is a free parameter
in our problem. In our non-dimensionlization, this value
is fixed by utilizing the 
observational evidence that the particle diffusivity of light elements
(lithium) is about $D_T \sim 10^3$ cm$^2$ s$^{-1}$ to be consistent
with the present solar surface lithium depletion
(e.g., see Barnes, Charbonneau \& MacGregor 1999).
Since $\eta_T \sim (\eta/D) D_T$ where
$D \sim 20$ cm$^2$ s$^{-1}$ and $\eta \sim 10^4$ cm$^2$ s$^{-1}$,
we impose the condition that $\eta_T
\sim 10^5$ cm$^2$ s$^{-1}$ for the parameter values typical of
the present sun (i.e. $B_y \sim 10^5$ G, 
$\OOmega \sim 3 \times 10^{-6}$ s$^{-1}$, etc).
[Note that $\eta_T$ in Eq. (\ref{eq30})--(\ref{eq31}) depends
on both $\OOmega$ and $B_y$, which change in time according to
Eq. (\ref{eq40}).] 

%
   \begin{figure}
   \centering
   \includegraphics[width=10cm]{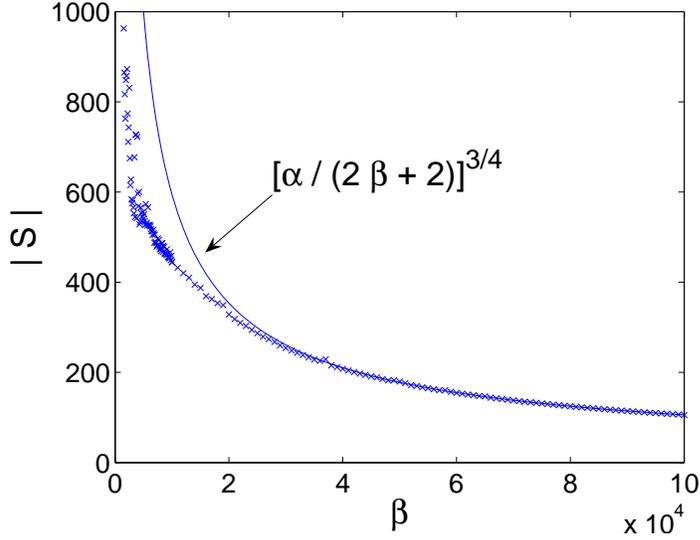}
      \caption{The plot of the amplitude of $\mA = \OOmega/\Omega$
               as a function of $\beta$ ($\propto$ hyperviscosity) 
               for $\alpha = 10^8$. The cross 
               symbols represent the results of the numerical simulation
               of the toy model Eq. (\ref{eq40}). The dashed line
               denotes the asymptotic behavior, given by
               $[\alpha/(2 \beta + 2)]^{3/4}$. 
              }
         \label{Fig1}
   \end{figure}

The results from the numerical simulation of Eq. (\ref{eq40}) 
for $\alpha = 10^8$ are plotted by the cross symbols in Fig. 1, which shows
$|\mA| = |\OOmega/\Omega|$ for various values of $\beta$.
The results nicely show that the amplitude of
$\mA$ decreases as $\beta$ (hyperviscosity) becomes large,
as expected.
However, for small $\beta$ (hyperviscosity),
an amplitude of $\OOmega > 10^3 \Omega$ is too large to be
reasonable. For this large value of $\OOmega$, 
which is comparable to the Brunt-V\"ais\"al\"a frequency $N$, 
the shear flow
is likely to be unstable against a shear instability
(see, e.g., Drazin 1981).
As $\beta$ increases, the amplitude of $\OOmega$ decreases
and approaches the asymptotic value given by the dashed line
$[\alpha/(2 \beta + 2)]^{3/4}$, which is obtained by
using $|\mA|=|\mB|$. The asymptotic value
indicates that $\OOmega \sim 100 \Omega$ for $\beta = 10^5$
while $\OOmega \sim 3\Omega$ for $\beta = 10^7$.
By using the definition of $\beta$, one can easily show that 
a reasonable value of shear $\OOmega \sim \Omega$
is possible only for an extremely thin tachocline with $h/\R \sim 10^{-5}$.
Since $|\mA|=|\mB|$ along the dashed line in Fig. 1, 
the toroidal magnetic field also seems too strong
($> 10^5$ G) to be stable (e.g. against the magnetic
buoyancy instability). 
To summarize, the results from our toy model (\ref{eq40}) clearly 
demonstrate
that for reasonable parameter
values, a stable stationary Hartmann layer is very unlikely
in the tachocline. 
Thus, a large-scale shear flow is likely to
become time-dependent (similarly to the behaviour found
in Kim and MacGregor 2001), or to develop
a secondary instability.
%

\section{In the limit of the strong radiative damping}
The anti-diffusive momentum transport discussed in the previous
sections originates from 
the stable stratification in the tachocline. 
The buoyancy force can however be weakened by
a strong radiative damping, as demonstrated in
Sect. 2.
Thus, it is conceivable that if a radiative damping
is large enough to weaken the buoyancy force sufficiently, 
the momentum transport may become diffusive, smoothing
out the gradient of the radial differential rotation.
In this section, we explore this possibility by
considering the limit of a strong radiative damping
where the density fluctuation $\rho_1$ in Eq.\ (\ref{eq3})
is stationary with $(\p_t + U_0 \p_y) \rho_1 = 0$.
In order for this limit to be valid, the parameter $\xi_\mu
= \mu k_y^2/\OOmega$, which has been assumed to be small in
the previous sections,
is no longer small. We are however still interested in the
case where the dissipation rate of magnetic field is small
compared to the shearing, namely, for $\xi_\eta = \eta k_y^2/\OOmega
\ll 1$. Note that this is a valid assumption for the reasonable
values of the characteristic scale of the forcing $l_f = 1/k_y > 10^5$ cm
in the tachocline.

By using $N^2 v_x/g + \mu \nabla^2 \rho_1=0$, the coupled
equations (\ref{eq1})-(\ref{eq2}) can easily be combined to yield 
\begin{equation}
\left [\p_\tau + {\mN^2 \over (1+\tau^2)^2} \right]
[(1 +\tau^2)\p_\tau \ha] + {\gamma^2} (1+\tau^2) \ha
\simeq  {i B_0 \over \OOmega^2 k_y} {\hf}(\tau)\,,
\label{eq81}
\end{equation}
to leading order in $\xi_\eta \ll 1$.
Here, ${\hat a} = {\tilde a} \exp{[\eta (k_x^3/3 k_y \OOmega + k_y^2 t)/2]}$;
${\hf} = {\tilde f} \exp{[\eta (k_x^3/3 k_y \OOmega + k_y^2 t)/2]}$;
$\tau = k_x/k_y = k_x(0)/k_y + \OOmega t$, $\gamma = |B_0 k_y/\OOmega|$,
$\omega_A = |B_0 k_y|$, and $\mN^2 = N^2/\OOmega (\mu k_y^2)$. 
The solution to Eq. (\ref{eq81}) can be obtained in the limit $\xi_\eta < 1$: 
\begin{equation}
\ha (\tau) \simeq {i B_0 \over \OOmega^2 k_y \gamma}
\int d\tau_1 {\sin{\gamma[\vphi(\tau)-\vphi(\tau_1)]}\hf(\tau_1)
e^{-\mN^2(\vpsi(\tau)-\vpsi(\tau_1)}
\over (1+\tau^2)^{1/2} ({1+\tau_1^2})^{1/2} \psi(\tau_1) }\,.
\label{eq82}
\end{equation}
Here, $\vphi(\tau)= \int^\tau d\tau_1 \psi (\tau_1)$,
$\psi(\tau) = 1 - [(1+\tau^2)^{-2} - \mN^2\tau(1+\tau^2)^{-3}
+ \mN^4(1+\tau^2)^{-4}/4]/2\gamma^2$, and
$\vpsi(\tau) = [\tan^{-1}{\tau} + \tau/(1+\tau^2)^{}]/4$.
By omitting the details of the algebra, we here
provide the resulting eddy viscosity $\nu_T$,
magnetic diffusivity $\eta_T$ and heat diffusivity $\mu_T$:
\begin{eqnarray}
\nu_T 
&\simeq& {\tau_f\over  4 B_0^2} \int {d^2k \over (2 \pi)^2}
{\phi(\bk)\over k_y ^4} \left [1-G(\mN^2)\right] 
\sim   {1 \over \omega_A^2} {\cal F}\,,
\label{eq90}\\
\eta_T 
&\simeq& {\tau_f\over  2 B_0^2} \int {d^2k \over (2 \pi)^2}
{\phi(\bk)\over k_y ^4} \left ({\xi_\eta \over 3}\right)^{2/3}
\Gamma \left({1\over 3}\right) e^{-\pi\mN^2/4} 
\lapprox   {\xi_\eta^{2/3} \over \omega_A^2} {\cal F}\,,
\label{eq91}\\
\mu_T 
&\simeq& {\tau_f\over  \mN^2} \int {d^2k \over (2 \pi)^2}
{\phi(\bk)\over k_y ^2} 
\sim   {1 \over N^2} {\cal F}\,.
\label{eq92}
\end{eqnarray}
Here, $\omega_A = |k_y B_0|$, $k_y$ is the typical wavenumber 
of the forcing, and ${\cal F} \sim \tau_f \int d^2 k \phi({\bf k})/k_y^2
\sim v_f^2/\tau_f$ is the strength
of the forcing with a characteristic forcing velocity
$v_f$ and correlation time $\tau_f$.
The function $G(\mN^2)$ depends on 
$\mN^2$, taking at most a maximum value $\sim 0.37$ for ${\mN}^2
\sim 4 $ and vanishing as either ${\mN} \to 0$ or $\mN \to \infty$.
Since $G(\mN^2)<1$, the eddy viscosity in Eq. (\ref{eq90}) is now
positive. That is, 
turbulent momentum transport in the limit of  a strong 
radiative damping is diffusive, in contrast to the case
of weak radiative damping considered in Sec. 3. This can be shown to be
related to the equipartition between kinetic and magnetic energy
to leading order, in contrast to the weakly damped case in Sect. 3
where the excess of kinetic energy over magnetic energy causes
a negative viscosity. As expected, this is because the buoyancy
force is greatly reduced by strong radiative damping.
The results in Eqs. (\ref{eq90})--(\ref{eq92}) show that
both $\nu_T$ and $\eta_T$ are quenched largely by magnetic fields while
$\eta_T$ is also reduced by shear ($\propto \xi_\eta^{2/3}
\propto \OOmega^{-2/3}$). 
In comparison, the heat diffusivity in Eq. (\ref{eq92}) exhibits
a different behavior, being mainly inhibited by
buoyancy force. This is because the coupling of the density fluctuation 
to the fluid is feeble due to strong radiative damping. 

The comparison of Eqs. (\ref{eq90}) and (\ref{eq91}) reveals that
the magnetic diffusivity is smaller than eddy viscosity by a factor
of $\xi_\eta^{2/3} \ll 1$ (recall $\xi_\eta= \eta k_y^2/\OOmega$ is
a small parameter characterizing the strong shear limit, where $l_f
= 1/k_y$ is the characteristic scale of the forcing). The ratio
$\eta_T/\nu_T \ll 1$ is however larger than the value in the case of
weak damping in Sect. 3. This is an interesting result
since this ratio is crucial in the estimate of the tachocline thickness,
which has been set by the molecular values  with $\eta/\nu \sim 10^2$
in previous works. We emphasize that our predicted value
$\eta_T/\nu_T$ ($ \ll 1$) is much smaller that $\eta/\nu$ based
on  molecular values.  
Furthermore, the ratio of the effect of the hyperviscosity to that
of eddy viscosity on the shear flow is now roughly given
by $(B_0/\R N)^2( \R^2 k_y/h)^2 (\Omega/N)^2 \gapprox 1$
for reasonable parameter values.
That is, the effect of hyperviscosity due to meridional circulation
is crucial in maintaining the momentum balance in the tachocline.

Based on these observations, we now seek to obtain the estimate on the 
thickness of the tachocline
and the strength $B_y$ and $B_z$ 
by requiring $\p_t = 0$ and by ignoring $\nu_T$
in Eq.\ (\ref{eq38}).
A simple analysis of
Eqs.\ (\ref{eq38})-(\ref{eq39}),
with the help of Eqs. (\ref{eq91}) and (\ref{eq92}) 
and $U_y \sim \OOmega \R$, gives us
\begin{eqnarray}
{h/\R \over \R k_y} &\sim & {1\over \Lambda} \left({\OOmega\over N}
\right)^2 \xi_\eta^{-1/3}\,,
\label{eq93}\\
{B_z \over B_y} &\sim &{\eta_T \over \OOmega h^2} \,,
\label{eq94}
\end{eqnarray}
where $\xi_\eta = \eta k_y^2/\OOmega$ and $l_f= 1/k_y$ is 
the characteristic length scale of the forcing.
In a sharp contrast to the  previous works which use the molecular
values for the viscosity and magnetic diffusivity,
the result (\ref{eq93}) shows that the
tachocline thickness is independent of the strength of
magnetic fields. For instance,
for $k_y = 10^{-7} \sim 10^{-6}$ cm$^{-1}$, $h/\R = 0.01 \sim 0.05$
(for $\OOmega \sim 3 \times 10^{-6}$ s$^{-1}$, 
$N \sim 10^{-3}$ s$^{-1}$, and $\Lambda \sim 3$).
This estimate is much larger than $h/\R\sim 10^{-3}$
by Gough \& McIntyre (1998),
and is mainly due to the small magnetic diffusivity
with the value $\eta_T/\mu_T \sim \xi_\eta^{2/3} (N/\omega_A)^2
\ll 1$. That is, a more efficient momentum transport than
magnetic dissipation causes a thicker tachocline.
Equation (\ref{eq93}) implies that the tachocline thickness 
decreases as the stratification becomes stronger, which could
have some implications for other stars. 
On the other hand, Eq.\ (\ref{eq94}) shows that
the ratio of magnetic field strengths depends
on the tachocline thickness and the turbulent magnetic diffusivity
$\eta_T$. 
As $\eta_T$ is directly proportional to the strength of the
forcing ${\cal F}$, we again utilize the constraint on 
the turbulent particle diffusivity $D_T \sim 10^3$ cm$^2$ s$^{-1}$
in the tachocline to be consistent with the surface depletion 
of lithium on the Sun. Since $\eta_T \sim (\eta/D) D_T$ (see
Kim 2006) where
$D \sim 20$ cm$^2$ s$^{-1}$ and $\eta \sim 10^4$ cm$^2$ s$^{-1}$,
we obtain $\eta_T \sim 10^5$ cm$^2$ s$^{-1}$. By using
this value in Eq. (\ref{eq94}) and $h/\R \gapprox 0.01 \sim 0.05$, we can obtain
$B_z/B_y \sim 10^{-8} \sim 10^{-7}$.
If we take, $B_y \sim 10^4$ G, $B_z \sim 10^{-4} \sim 10^{-3}$ G
while for $B_y \sim 10^5$ G, $B_z \sim 10^{-3} \sim 10^{-2}$ G.
These values of $B_z$ are comparable to the previously estimated
value. It is important to note that 
for $\eta_T \sim 10^5$ cm$^2$ s$^{-1}$, Eqs. (\ref{eq90})
and (\ref{eq92}) give 
$\nu_T \sim 10^6$ cm$^2$ s$^{-1}$ 
and $\mu_T \sim 10^{10}$ cm$^2$ s$^{-1}$ ($\gg \mu$).
Therefore, our results naturally predict a more efficient
momentum transport than particle transport (recall $D_T \sim 10^3$ cm$^2$ s$^{-1}$)
in the tachocline without fine-tuning parameters (see also Kim 2006),
as required to be consistent with the observations.

\section{Discussion and conclusions}
We have presented a theory of turbulent transport
in the sheared stratified magnetised tachocline 
when turbulence is driven and maintained by an external
forcing (e.g. due to plumes penetrating from the convection zone).
Despite the presence of a strong (uniform) magnetic field,
the momentum transport is found to be anti-diffusive ($\nu_T<0$)
for reasonable parameter values in the tachocline.
The magnetic diffusivity ($\eta_T$) and heat
diffusivity ($\mu_T$) are found to be positive, but severely reduced 
as a result of the shear stabilization and transport 
reduction by Alfven-gravity waves with very small values
of $\eta_T/|\nu_T|$ ($\ll 1$) and $\mu_T/|\nu_T|$ ($\ll 1$. 
Since the momentum transport is anti-diffusive, the
gradient in the radial differential rotation is
amplified. As the shear becomes strong, the
hyperviscosity due to the meridional circulation, which
is always diffusive, can become
important, counteracting the effect of
negative eddy viscosity. However, in order to maintain
the radial shear $\OOmega$ in the differential rotation
with a value comparable to the average solar rotation
$\Omega$, as presently inferred, the tachocline as the Hartmann layer
has to be very thin with the
thickness $h/\R \sim 10^{-5}$ for the toroidal magnetic field
$B_y \sim 10^5$ G. Here, $\R$ is the solar radius.
Otherwise, the radial shear and/or toroidal 
magnetic field in the tachocline appears to be too
strong to be stable against the shear instability
and/or magnetic buoyancy instability. 
Furthermore, since the turbulent coefficients
depend nonlinearly on $\OOmega$ and $B_y$, 
$\OOmega$ and $B_y$ form a nonlinear dynamical system,
which can 
exhibit complex time variation. 
These results point to the possibility
that the tachocline may exhibit a much more complex (temporal
and/or spatial dependent) dynamics than previously thought 
even in the slow tachocline scenario. 

In the limit of strong radiative damping
where the temperature (density) fluctuation is almost
stationary, the turbulent momentum transport is found to 
be diffusive, down the gradient. This requires 
the turbulence to be on very small scales such that 
$\xi_\mu = \mu k_y^2/\OOmega \gg 1$.
Here, $1/k_y=l_f$ is the characteristic length scale of
the forcing, and $\OOmega \sim 3 \times 10^{-6}$ s$^{-1}$
is the radial shear. 
Thus, the forcing scale $l_f$ has to be smaller than $10^6 \sim 10^7$ cm,
which is about $10^{-4} \R$. In this case, the tachocline
may be viewed as a Hartmann layer with possibly
stable configurations of the radial differential rotation
and toroidal magnetic field. A simple analysis by using
the predicted values of $\nu_T$, $\mu_T$, and $\eta_T$
suggested that the effect of the hyperviscosity is likely
to dominate over the eddy viscosity. However, since the magnetic 
diffusion is severely quenched compared with heat diffusivity,
the momentum transport is effectively much more efficient than
the magnetic field diffusion. Consequently, 
the major force
balance in the tachocline leads to the estimate of
the tachocline thickness ($\gapprox 10^{-2} \R$),
which is larger than the
previously estimated value  $\sim 10^{-3} \R$ by Gough \& McIntyre 
(1998).
As $\eta_T$ and $\mu_T$ depend on
the strength of toroidal magnetic fields and radial shear,
the tachocline thickness is found to be independent of
the strength of (toroidal and/or poloidal) magnetic field
(c.f. Gough \& McIntyre 1998).
The strength of poloidal magnetic field 
in the radiative interior is estimated to be of order
$10^{-4} \sim 10^{-2}$ G for the value of the toroidal
magnetic field $10^4\sim 10^5$ G, comparable to
the previously estimated value.
Importantly, the results provide a natural 
explanation for a more efficient momentum transport than particle
mixing in the tachocline. However, 
a severe reduction in the magnetic field
diffusion could be problematic for the solar
dynamo [e.g., the interface dynamo (Parker 1993)]. 

While the discussion of our results is focused
on the applications to the present sun,
they might also have interesting implications for other
stars. In particular, our predicted values of turbulent
transport coefficients have different dependences
on the strength of magnetic field, stratification, and
radial shear as well as on the molecular values of
viscosity, ohmic diffusivity, and radiative diffusivity.
As the values of these parameters vary from one star to another,
it would be of interest to explore the implications of
these results for other stars.
For example, it might be possible to utilize
the results to infer the value of the radial shear or the strength
of magnetic field in other stars, and also to gain some
insight into the presence of a tachocline-like shear layer in those.
Of course, in the case of more massive stars, the 
rotation rate is much faster the sun, demanding the prediction
of turbulent transport coefficients by taking into account
the effect of average rotation (Leprovost \& Kim 2006). 
Ultimately, it will be interesting to investigate
a consistent model incorporating the spin-down
of the sun (and other stars) due to the angular momentum loss.
Finally, we note that our analysis, limited to 2D, should be
extended to 3D, in particular, to study solar dynamos (e.g. the $\alpha$
effect).
These issues are currently under investigation 
and will be addressed in future publications.

\begin{acknowledgements}
This work was supported by the PPARC Grant PP/B501512/1.
\end{acknowledgements}


%
%

\end{document}